\documentclass[conference]{IEEEtran}

\usepackage{amsmath}
\usepackage{graphics}
\usepackage{setspace}
\usepackage{cite}
\usepackage{latexsym}
\usepackage{float}
\usepackage{epsfig}
\usepackage{multirow}
\usepackage{cite,cases,url}
\usepackage{amssymb}
\usepackage{graphicx}
\usepackage{epstopdf}
\usepackage{balance}
\usepackage{enumerate}
\usepackage{array}
\usepackage{algorithmic}
\usepackage{algorithm}

\newcolumntype{L}{>{\centering\arraybackslash}m{5cm}}
\newcolumntype{K}{>{\centering\arraybackslash}m{6cm}}
\newcolumntype{P}{>{\centering\arraybackslash}m{2.3cm}}
\newcolumntype{M}{>{\raggedright\arraybackslash}m{2cm}}
\newcolumntype{N}{>{\raggedright\arraybackslash}m{2.5cm}}

\begin{document}

\title{Security and Protocol Exploit Analysis of the 5G Specifications}

\author{\IEEEauthorblockN{Roger Piqueras Jover}
\IEEEauthorblockA{Bloomberg LP\\
New York, NY\\
rpiquerasjov@bloomberg.net}
\and
\IEEEauthorblockN{Vuk Marojevic}
\IEEEauthorblockA{Dept. Electrical and Computer Engineering\\ Mississippi State University,
Mississippi State, MS\\
vuk.marojevic@msstate.edu}}
\maketitle


\begin{abstract}

The Third Generation Partnership Project (3GPP) released its first 5G security specifications in March 2018. This paper reviews the proposed security architecture, its main requirements and procedures, and evaluates them in the context of known and new protocol exploits.
Although security has been improved from previous generations, 
our analysis identifies unrealistic 5G system assumptions and protocol edge cases that can render 5G communication systems vulnerable to adversarial attacks. For example, null encryption and null authentication are still supported and can be used in valid system configurations. With no clear proposal to tackle pre-authentication messages, mobile devices continue to implicitly trust any serving network, which may or may not enforce a number of optional security features, or which may not be legitimate. Moreover, several critical security and key management functions are left outside of the scope of the specifications. 
The comparison with known 4G Long-Term Evolution (LTE) protocol exploits reveals that the 5G security specifications, as of Release 15, Version 1.0.0, do not fully address the user privacy and network availability challenges. 


\end{abstract}

\IEEEpeerreviewmaketitle

\section{Introduction}
\label{sec:intro}
The Third Generation Partnership Project (3GPP) published its fifteenth release of the mobile communication system specifications in March 2018, setting the foundations for the 5th generation of mobile communication (5G). With ground breaking upgrades at the radio layer, the New Radio (NR) standard implements an advanced physical layer that supports millimeter wave communications and antenna arrays for massive multiple-input, multiple-output (MIMO) systems \cite{swindlehurst2014millimeter}. In parallel, the 5G core network (5GC) has been redesigned for enhanced flexibility and service versatility. 
The goal of 5G networks is to provide ubiquitous, high-speed, and low-latency connectivity for enhanced mobile broadband, massive machine type communication and real-time control. 
5G will enable the tactile Internet, untethered augmented and virtual reality, smart connected vehicles and further new connectivity types \cite{docomo2014docomo,boccardi2014five}.

As with its preceding generations--2G, 3G and the 4G Long Term Evolution (LTE)--, security is of capital importance for 5G networks and services. Cellular communication networks provide connectivity to billions of civilians worldwide. They are also the connectivity cornerstone for current and emerging critical infrastructure, supporting the smart grid, first responder units, and advanced military operations. The advent of 5G will enable new verticals in the civilian, industrial and mission-critical domains \cite{boccardi2014five}. 

Motivated by the inherent security weaknesses of legacy 2G networks, such as the lack of mutual authentication between the network and the user equipment (UE), security has been one of the key design considerations for mobile communications starting with 3G. LTE implements strong encryption and integrity protection algorithms, backed with a mutual authentication using symmetric keys that are securely stored in the Universal Subscriber Identification Module (USIM) and the operator's Home Subscriber Server (HSS) \cite{LTE_SECURITY_3GPP}. Nevertheless, 
a series of vulnerabilities inherent to the LTE protocol still exist and have been identified by researchers over the last few years. For example, a substantial number of pre-authentication messages are sent in the clear, which can be exploited to launch Denial of Service (DoS) attacks and obtain location information of mobile subscribers~\cite{rupprecht-19-layer-two,jover2016exploits,LTEpracticalattacks}.

The first release of the LTE specifications, 3GPP Release 8, was published in 2007. The main security vulnerabilities were not identified and reported in open literature until much later, though. One of the reasons for this was the lack of available and affordable tools for LTE security research. LTE open-source software libraries running on personal computers and using commercial off-the-shelf software-defined radio (SDR) peripherals did not reach a sufficient level of maturity until recent years. Once they became available, a wave of excellent security research in the area of LTE mobile communications emerged and identified numerous protocol vulnerabilities \cite{jover2016opensource,RaoMILCOM17,rupprecht-19-layer-two,hussain2018lteinspector,LTEpracticalattacks}.

As in LTE, security is a key consideration and core aspect for the definition and specification of 5G systems. Since the inception of the communication protocols for NR and 5G-S (5G System), there has been a substantial effort in addressing known LTE protocol exploits, with particular focus on preventing International Mobile Subscriber Identifier (IMSI) catchers or Stingrays \cite{IMSIcatcher}. As a result, the 5G protocol introduces 
the Subscriber Permanent Identifier (SUPI), as replacement of the IMSI, and a Public Key Infrastructure (PKI), which allows the encryption of the SUPI into the Subscriber Concealed Identifier (SUCI) \cite{5G_SECURITY_3GPP}. 


Preventing protocol exploits that leverage pre-authentication messages was also a key security design goal for 5G \cite{LTE_SECURITY_STUDY}. Nevertheless, and despite the efforts to design a secure architecture, a number of insecure protocol edge cases 
still exists. Moreover, there is no clear solution yet to prevent the implicit trust of pre-authentication messages, which can be exploited by an adversary to both deny the service to subscribers as well as intercept sensitive user information \cite{LTEpracticalattacks}. 

If the PKI architecture that is used to conceal the SUPI 
is also intended to prevent other protocol exploits, full security against such exploits can only be achieved if all USIMs in all mobile devices had the public keys of all operators in the world. In addition, all operators would need to keep the corresponding private keys well secured. Not only such key management and rotation is unfeasible and, as of Release 15, left outside the standard specifications, but political and operator disagreements would most likely result in the lack of global adoption. Insecure protocol implementation and exploitation of pre-authentication messages could be the consequences.

This paper provides a wide-angle analysis of the 5G access network security architecture and procedures and its potential deployment challenges as a result of the security framework described in 3GPP TS 33.501~\cite{5G_SECURITY_3GPP}. 
The underlying requirements and assumptions for 5G security are identified and analyzed holistically, with specific focus on global adoption and the resulting consequences. The objective of this paper is not to provide a comprehensive analysis of the security of 5G network elements and layers, but rather to globally assess the critical security challenges 
of the current 5G security specifications with an outlook at future 5G network deployments. 

The remainder of this paper is organized as follows. Section \ref{sec:arch} provides an overview of the 5G security architecture and components, setting the context for Section \ref{sec:requ}, which discusses the main 5G security requirements and procedures of the first 3GPP Release 15 specifications. Section \ref{sec:vuln} provides a holistic analysis of the deployment challenges of the proposed 5G security framework, highlighting the potential risk of protocol exploits and sensitive information leaks. 
The 5G security framework is then analyzed in terms of the known LTE protocol exploits in Section \ref{sec:lte}. Finally, Section \ref{sec:conclusions} summarizes our findings and proposes research directions to address the identified security vulnerabilities.

\section{Overview of the 5G security architecture}
\label{sec:arch}
The 5G security architecture spans across the UE, radio access network, core network and application \cite{5G_SECURITY_3GPP}. The architecture is correspondingly organized into an application stratum, a serving stratum, and a transport stratum. Figure \ref{arch} shows a simplified diagram of the serving stratum and the transport stratum. Different security features are defined across the network and end user components, which combined create a secure system design:

\begin{figure}[h]
\begin{center}
\includegraphics[totalheight=0.16\textheight]{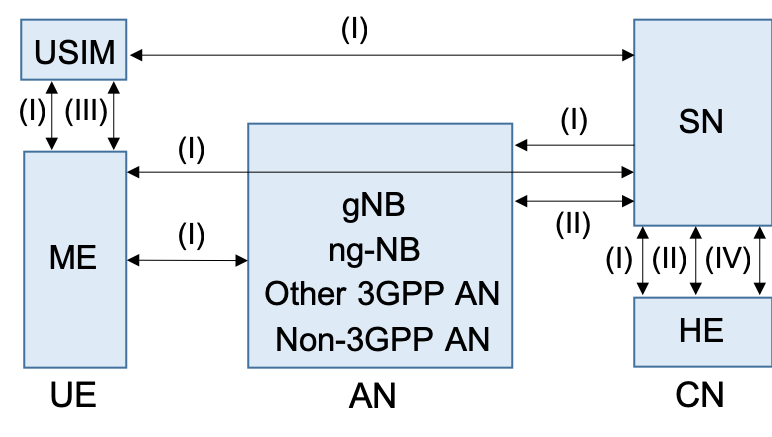}
\caption{5G security architecture (AN--Access Network, HE--Home Environment, ME--Mobile Equipment, SN--Serving Network).}
\label{arch}
\end{center}
\end{figure}

\begin{itemize}
\item Network access security (I): A set of features and mechanisms that enable a UE to authenticate and securely access network services. UEs therefore exchange protocol messages through the access network with the serving network (SN) and leverage the PKI, where keys are stored in the USIM and the home environment (HE).
\item Network domain security (II): A set of features and mechanisms that enable network nodes to securely exchange control plane and user plane data within 3GPP networks and across networks.
\item User domain security (III): A set of features and mechanisms at the UE that secure the access to mobile equipment and mobile services. It establishes hardware security mechanisms to prevent the mobile terminals and USIMs from being altered.
\item Service-Based Architecture (SBA) domain security (IV): A set of network features and mechanisms for network element registration, discovery and authorization, 
as well as for protecting the service-based interfaces. It allows new 5GC functions, which may be implemented as virtual network functions, to be securely integrated. It also enables secure roaming, which involves the SN as well as the home network (HN)/HE.
\item Visibility and configurability of security (not shown in Fig. \ref{arch}): A set of features and mechanisms that allow informing users whether a security feature is in operation. It can also be used to configure security features. The 3GPP 5G security specifications formally establish optional security features and degrees of freedom for implementation and operation, which means that 5G users may encounter different security context.
\end{itemize}

\begin{figure}[h!]
\begin{center}
\includegraphics[totalheight=0.25\textheight]{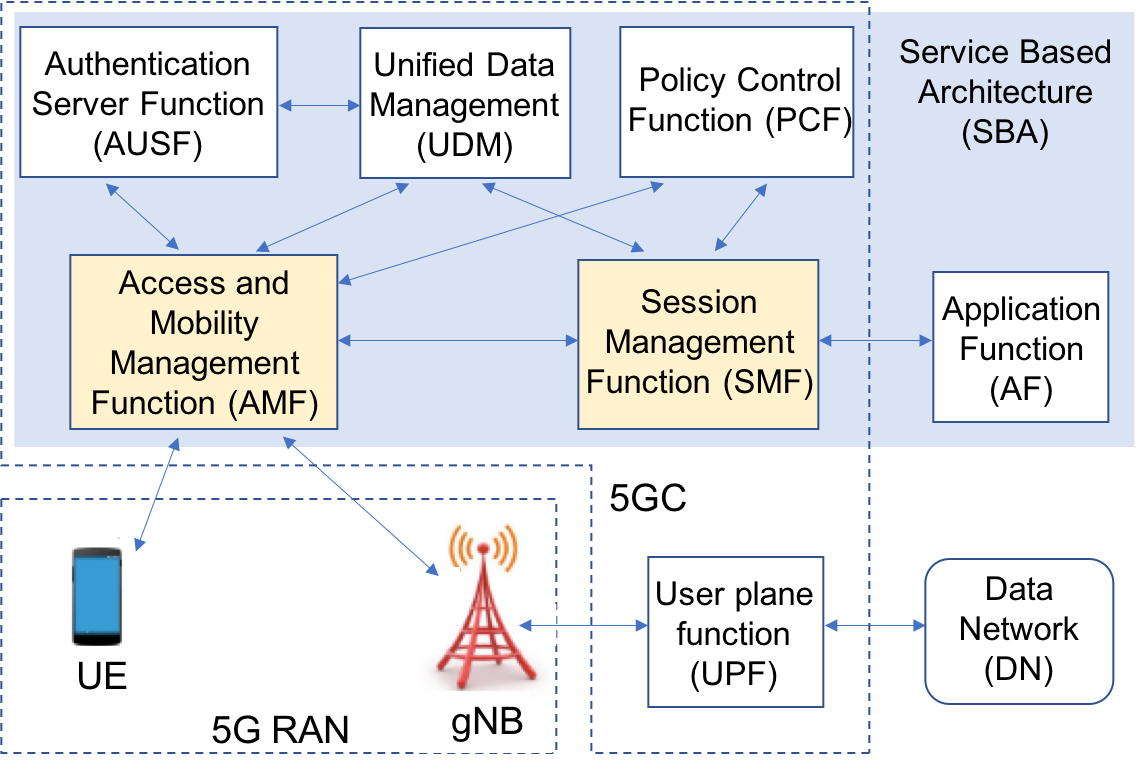}
\caption{Simplified 5G reference network architecture.}
\label{functions}
\end{center}
\vspace{-.1in}
\end{figure}

The 5G specifications define a number of network functions and their interfaces, enabling the data flow between the 5G radio access network (RAN), 5GC, and external networks. Figure \ref{functions} illustrates a simplified 5G reference network architecture. The network functions and security features specify a flexible, yet secure design for developing 5G mobile communication systems.

\section{5G security requirements and procedures}
\label{sec:requ}
The 5G security framework defines a series of security requirements, features and procedures \cite{5G_SECURITY_3GPP}, which we 
summarize in continuation. Table I captures the core security requirements and the corresponding procedures for the 5G RAN. This table highlights, in italics, some of the requirements and procedures that can lead to security vulnerabilities. These vulnerabilities and their potential implications are analyzed in subsequent sections. 

\begin{table*}[t]
    \begin{center}
    \caption {5G RAN security requirements and procedures (security domain association according to Fig. 1).}
    \begin{tabular}{|p{1.1cm}|p{11cm}|p{4.5cm}|}
    \hline
    \textbf{Scope} & \textbf{Security requirements} & \textbf{Procedures}\\
    \hline
    \vspace{1pt}
    General (I) &
\begin{itemize}
\item Mitigation of bidding down attacks.
\item Mutual authentication.
\item UE, access and serving network authorization.
\item \textit{{Allowance for unauthenticated emergency services.}}
\end{itemize}
	&
    \vspace{1pt}
    Authentication procedures using EAP-AKA and 5G AKA authentication methods. \\
    \hline
    \vspace{1.5pt}
    UE and gNB (I) &
    \begin{itemize}
\item User and signaling data confidentiality protection through cyphering.  \textit{gNB triggered, considering UE security capabilities and SN's list of security capabilities. Null encryption supported. Confidentiality protection optional to use.}
\item User and signaling data integrity and replay protection. gNB triggered, \textit{considering UE security capabilities and SN's list of security capabilities. Null integrity protection supported. Integrity protection of user data optional to use.} RRC and NAS signaling protection mandatory, \textit{but exceptions exists, including unauthenticated emergency sessions.}
\end{itemize}
    &
    \vspace{1pt}
    Key derivation, distribution and agreements from a key hierarchy, supporting 128 bit key and 256 bit key encryption. For every key in a network entity, there is a corresponding key in the UE, with the root key stored in the USIM.  \\
    \hline
    \vspace{1pt}
    UE (III)&
\begin{itemize}
\item Secure storage and processing of subscription credentials using a tamper resistant secure hardware component.
\item Subscriber privacy through use of temporary and concealed subscriber identities (5G-GUTI and SUCI).
\item  \textit{If provisioned by the home operator, the USIM shall store the HN public key used for concealing the SUPI.}
\end{itemize} 
    &
    \vspace{1pt}
   \textit{{Null-scheme supported and shall be used when public key 
not provisioned by HN, which controls subscriber privacy and the provisioning and updating of keys.}}\\
    \hline
    \vspace{1pt}
    gNB (I), (II)&
\begin{itemize}
\item Authorized setup and configuration by O\&M through certificates, the use of which is optional.
\item Key management, \textit{optional for the 5G PKI-based architecture}.
\item Secure environments for keys, user plane and control plane data storage and processing.
\end{itemize}
    &
    \vspace{1pt}
    \textit{Authentication and key derivation may be initiated by the network as often as the operator decides when an active NAS connection exists.} \\
    \hline
    \end{tabular}
    \end{center}
    \label{tabReq}
    \vspace{-.1in}
\end{table*}

\subsection{Key Framework}
The 5G security procedures leverage a hierarchical key derivation, distribution, and management framework. Keys are stored in a number of network entities. The long term key \emph{K} is stored by the Authentication Credential Repository and Processing Function (ARPF) of the Unified Data Management (UDM) layer and the USIM holds the user corresponding copy of that symmetric key. All other keys are derived from it. The key generation and distribution is detailed in \cite{5G_SECURITY_3GPP}.

\subsection{Authentication and Home Control}
3GPP establishes the Extensible Authentication Protocol for Authentication and Key Agreement (EAP-AKA) and 5G AKA as the authentication methods that must be supported by both the 5G UEs and the 5GC. 5G UEs use the SUCI in their registration requests to initiate the authentication process using the method they select. These security modes are used for mutual authentication and subsequent service security and encryption procedures.  
5G AKA enhances the AKA protocol of 4G LTE \cite{LTE_SECURITY_3GPP} by providing the HN with proof of successful authentication of the UE from the visited network. 

\subsection{Security Contexts}
The 5G security specifications define a number of security contexts for different scenarios: a single 5G serving network (SN), across multiple SNs, and between 5G and 4G networks. When a UE is registered with two SNs, both networks must independently maintain and use a separate security context. When the UE is registered to two SNs in the same public land mobile network (PLMN), 3GPP and non-3GPP, the UE establishes two independent Non-Access Stratum (NAS) plane connections with those networks, but uses a common NAS security context consisting of a single set of keys and security algorithms.

\subsection{State Transition and Mobility}
Procedures for maintaining or disregarding a security context during state transition and handover are also defined, to some extent, in \cite{5G_SECURITY_3GPP}. The specifications state that it is up to the operator's policy how to configure the selection of handover types. 
This decision is a function of the operator's security requirements, thus leaving the security during handovers as an opt-in feature instead of enforcing it through the standard. As a consequence, an operator could potentially implement an insecure handover procedure.

\subsection{Non-Access Stratum}
Cryptographic separation and replay protection of two active NAS connections is supported through a common NAS security context, which has parameters that are specific to each NAS connection. NAS uses 128-bit ciphering algorithms for integrity and confidentiality protection. {Note that both null encryption and null integrity protection are supported. If the UE has no NAS security context, the initial NAS message is sent in the clear and contains the subscription identifier 
and UE security capabilities, among others.} 

\subsection{Radio Resource Control}
The Radio Resource Control (RRC) integrity and confidentiality protection is provided by the packet data convergence protocol (PDCP) layer between the UE and gNB and no layers below PDCP shall be integrity protected. Replay protection is to be activated when integrity protection is activated, {except when the selected integrity protection algorithm is NIA0 (null integrity protection).}
RRC integrity checks are performed both at the UE and the gNB. In the case where a failed integrity check 
is detected after the start of the integrity protection, the concerned message shall be discarded.

\subsection{User Plane}
The Session Management Function (SMF) shall provide user plane security policy for a protocol data unit (PDU) session to the gNB during the PDU session establishment procedure. {If user plane integrity protection is not activated for data radio bearers (DRBs), the gNB and the UE shall not integrity protect the traffic of such DRB. 
If user plane ciphering is not activated for DRBs, the gNB and the UE shall not cipher the traffic of such DRB. The local SMF can override the confidentiality option in the user plane security policy received from the home SMF.}

\subsection{Subscription ID Privacy}
The SUCI is the concealed version of the 5G permanent subscription identifier SUPI. The SUCI is transmitted over the air to prevent exposing the user identity in the clear. It is constructed from the SUPI using the operator's public key and a probabilistic asymmetric encryption method to prevent identity tracking. 
Nevertheless, the SUPI null protection scheme is used for unauthenticated emergency sessions, when so configured by the HN, or when the operator public key has not been provisioned.

The 5G specifications also define a temporary identifier, the 5G Global Unique Temporary Identifier (5G-GUTI), to minimize the exposure of the SUPI or SUCI. 5G-GUTI is to be reassigned based on UE triggers, but it is left to the implementation to determine the rate of such reassignment.


\section{Potential vulnerabilities of 5G---security challenges and opportunities}
\label{sec:vuln}
As introduced in Sections \ref{sec:arch} and \ref{sec:requ}, 5G mobile networks implement a security architecture similar to that of LTE systems, with a small  difference in how trust and security is established. 
Pre-5G communication systems base all security functions on symmetric keys that are securely stored both in the USIM and the HSS. Based on the shared secret key $k_s$, an LTE UE can authenticate the network and the network can authenticate the UE. The encryption and integrity protecting keys are derived from $k_s$ \cite{LTE_SECURITY_3GPP}. This symmetric key security architecture results in the inability of a communication endpoint, the UE, to verify the authenticity and validity of any message that is exchanged \emph{prior to} the \emph{NAS Attach} cryptographic handshake. The need for pre-authentication messages to be sent in the clear is widely acknowledged as the root cause of most known LTE protocol exploits \cite{jover2016exploits,LabCM2017,LTEpracticalattacks}.

\begin{table*}[t]
    \begin{center}
      \caption {Summary of security and implementation challenges of 3GPP 5G Release 15.}
        \begin{tabular}{|p{5.5cm}|p{5.5cm}|p{5.5cm}|}
    \hline
    \textbf{Security/ implementation challenge} & \textbf{Root cause} & \textbf{Impact}\\
    \hline
    PKI infrastructure & Considered out of scope of the 3GPP specifications & Implementation specific, potential for not being implemented \\
    \hline
    Key management (rotation, over-the-air provisioning, etc.) & Considered out of scope of the 3GPP specifications & Implementation specific, potential for not being implemented \\
    \hline
    Global cooperation & Security against pre-authentication message exploits guaranteed only if USIM contains a public key for every operator worldwide & With just one operator or country being non compliant, system security and user privacy can be compromised through rogue base stations and spoofed pre-authentication messages \\
    \hline
    Support for NULL encryption and NULL integrity & Requirements from standards stake holders and lawful interception working group & Potential for bidding down attacks and rogue base stations, especially if no public key provisioned for the operator \\
    \hline
    \end{tabular}
    \end{center}

    \label{tab:5Gsecuritychallenges}
\end{table*}

For any communication protocol, including 5G, independently of how strong a security architecture is and how sophisticated its cryptographic algorithms are, it only takes one single edge case or insecure function to defeat the entire system. For example, although in LTE the IMSI should only be sent in the clear the very first time a mobile phone is switched on, there is a number of legitimate and explicitly defined use cases in which the network can request that the UE identifies itself using its IMSI. 


Clear guidance through standardization and its enforcement are the basis for global compliance. Security functions and procedures that are left out of the scope of the protocol specifications can result in insecure edge cases. Therefore, critical security features and mechanisms cannot be optional and 
all operators need to opt-in for implementing these and implement them rigorously.

\subsection{Pre-Authentication Message Exploits}
The goal of the 5G security architecture is to 
tackle the challenge of pre-authentication messages and other protocol exploits 
\cite{LTE_SECURITY_STUDY}. By introducing the concept of operator public keys, 5G systems provide the tools for establishing a root of trust between the end user and the mobile operator under the umbrella of the 5G PKI. Leveraging the public keys burned into USIMs, operators can 
securely receive encrypted messages from the UEs 
as well as sign messages with their corresponding secret key to be validated by the UEs.

This PKI is the method that is proposed to protect against Stingrays. 
However, there is no clear solution in the specifications on how to achieve such level of security against all protocol exploits leveraging 5G pre-authentication messages. The specifications fall short of an ideal full PKI architecture, leveraging digital certificates and a Certificate Authority (CA), to tackle such a security challenge. Moreover, although 
SUPI catching is substantially more challenging than IMSI catching, there is still a number of valid protocol edge cases in which the SUPI is transmitted in the clear \cite{5G_SECURITY_3GPP}. Therefore, a rogue 5G base station could potentially trick a UE into disclosing its SUPI. 

It is worth noting that in 5G there is no method to prevent a rogue base station from instructing a UE to disclose its SUPI leveraging a spoofed pre-authentication message. 
However, in 5G, the SUPI would be transmitted encrypted in the form of the SUCI. 
Similarly, no security method in 5G prevents a UE from implicitly trusting pre-authentication messages.

In order to avoid pre-authentication message exploits using the 5G current PKI basic proposal, global compliance would be necessary. 
That is, in order to verify the validity of all pre-authentication messages in all connectivity scenarios, including roaming, each UE would require a cryptographic root of trust for any network it may connect to. This is so because network originating messages such as \textit{AttachReject} and \textit{TAUReject}, known for their LTE protocol exploits~\cite{jover2016exploits,LTEpracticalattacks}, could originate at the visiting network, as opposed to the HN, while roaming.

A potential solution against pre-authentication message exploits would also require operators to load into all USIMs and properly manage a public key for every operator in every country, without exception. It is anticipated that some countries will ban the public keys from certain other countries or operators, something that has been observed before 
\cite{HuaweiBan}.

In general, protocol exploits like the ones disclosed in~\cite{hussain2018lteinspector,jover2016lte,LTEpracticalattacks} are, as of Release 15, Version 1.0.0, still possible in 5G. 


\subsection{Other Security Challenges}

The new security framework and architecture is considered fundamental for securing emerging 5G mobile networks. However, our initial analysis of the security architecture already highlights a number of remaining security weaknesses that should be addressed. Table II 
identifies the key 5G security challenges, their root causes and potential impacts. 
The specifications leave outside of the scope of the protocol most implementation details that are critical for security, such as the key management of operator public keys residing in the subscribers' USIMs, the structure of certificates and how or whether keys are ever rotated \cite{5G_SECURITY_3GPP}. It is left for the industry to figure those details out. Prior experience has shown that rapid roll out and affordable service delivery require simple protocol solutions, which oftentimes compromise security \cite{LabSM2017}. In addition, lawful interception requirements mandate continuing support for null encryption and null integrity protection, which results in insecure modes of communication and protocol edge cases.


Increased home control with respect to prior generations is considered useful for preventing certain types of fraud. The proposed 5G security framework supports implementing such procedures, but they are considered beyond the scope of the specifications: \textit{the actions taken by the home network to link authentication confirmation (or the lack thereof) to subsequent procedures are subject to operator policy and are not standardized} \cite{5G_SECURITY_3GPP}.

In addition to the aforementioned security challenges, researchers are already finding weaknesses in the cryptographic operations defined in~\cite{5G_SECURITY_3GPP}. The authors of~\cite{5GAKAformalanalysis2018} use formal verification tools to analyze the 5G AKA algorithms and demonstrate that the protocol fails in meeting several security goals, which are explicitly required. The study also shows that the 5G protocol lacks other critical security properties. The studies presented in~\cite{5GAKAsecurity} and~\cite{AKAkoutsos20185g} reach similar conclusions. Moreover, the authors of~\cite{5Gdowngrade} describe  potential downgrade attacks against 5G networks.


\subsection{PKI-Based Architecture Alternative}

The move towards a PKI-based architecture in 5G is a step in the right direction. PKI systems provide a wider flexibility for sophisticated security solutions that could potentially tackle, for example, the challenge of pre-authentication messages. However, such a critical element of the 5G system architecture should not be left outside of the specifications.

Global agreement and adoption of a large scale PKI architecture is necessary for fully addressing the security challenges in 5G in the long term. However, instead of basing the system on public keys burned into the USIM, an improved architecture would include a global 5G Certificate Authority (CA). The CA would act as the root of trust to authenticate messages and communication based on digital certificates~\cite{housley1998internet}. Such authority would provide a more flexible architecture, and the corresponding certificate revocation and management challenges have already been addressed and the solutions vetted by the secure Internet implementation community~\cite{SSL}.

Similar proposals about the potential of PKI-based architectures applied to mobile communication systems have been discussed for over a decade now~\cite{KambourakisPKI}. It is also an important element of the European 5GPPP group~\cite{5GPPPcerts}.

\section{Impact of LTE protocol exploits on 5G}

\label{sec:lte}
The LTE security architecture was designed to address the challenges of previous generations. The first generation of mobile networks (1G) lacked support for encryption and this was one of the main drivers for the introduction of 2G digital mobile communications. Legacy 2G networks do not support mutual authentication and use an encryption algorithm that is outdated \cite{attack_sniffing}. LTE implements specific functionalities to guarantee the confidentiality and authenticity of mobile networks and messages, using much stronger cryptographic algorithms and explicit mutual authentication between the UE and the eNodeB. 
This makes 4G LTE inherently more secure than prior generations, yet still vulnerable to certain exploits.

\begin{table*}[th]
    \begin{center}
    \caption {Major LTE protocol exploits and their impact on 5G.}
     \begin{tabular}{|p{3.5cm}|p{4.5cm}|p{8.5cm}|}
    \hline
    \textbf{LTE protocol exploit} & \textbf{Threat} & \textbf{Impact on 5G}\\
    \hline
    IMSI catching & Privacy threat, location leaks, SS7 leaks, etc. \cite{LTEpracticalattacks,jover2016exploits,LTEpracticalattacks_Blackhat,hussain2018lteinspector} & Potential for IMSI/SUPI catching in some protocol edge cases, such as when an operator does not implement optional security features or when an unauthenticated emergency call is maliciously triggered. \\
    \hline
    Attach/ Tracking Area Update (TAU) request & DoS 
    \cite{LTEpracticalattacks,jover2016exploits,hussain2018lteinspector} & DoS of 5G mobile devices exploiting pre-authentication messages with rogue base station broadcasting a valid Mobile Country and Network Code (MCC-MNC) combination for network with no public key provisioned in the USIM. \\
    \hline
    Silent downgrade to GSM & Man in the middle attacks, phone call and SMS snooping \cite{LTEpracticalattacks,jover2016exploits,hussain2018lteinspector} & Silent downgrade to GSM exploiting pre-authentication messages with rogue base station broadcasting an MCC-MNC of a network with no public key provisioned in the USIM. \\
    \hline
    Location tracking with RNTI & Location leaks, traffic estimation, service estimation \cite{rupprecht-19-layer-two} & Potential device location traffic and traffic profiling \\
    \hline
    Insufficient protection of DNS traffic at layer 2 & DNS hijacking over LTE \cite{rupprecht-19-layer-two} & Man in the middle attacks, credential stealing, remote malware deployment. \\
    \hline
    \end{tabular}
    \end{center}
    \label{tab:LTEvs5G}
\end{table*}

\subsection{LTE Protocol Exploits}
The existence of LTE protocol vulnerabilities has been known for some time, although these have not been publicly discussed until recently. 
The openness of the standard, the large community of researchers, and the broad availability of SDRs, software libraries and open-source implementations of both the eNodeB and the UE protocol stacks have enabled a number of excellent LTE security analyses \cite{RaoMILCOM17,MarVTC17,LabICNC2015,rupprecht-19-layer-two,hussain2018lteinspector}. 
Despite the stronger cryptographic algorithms and mutual authentication, UEs and base stations exchange a substantial amount of pre-authentication messages that can be exploited to launch denial of service (DoS) attacks \cite{MINA_JP,LabCM2017,jover2016exploits}, catch IMSIs \cite{Nor2016} or downgrade the connection to an insecure GSM link \cite{LTEpracticalattacks,hussain2018lteinspector}. Researchers also found new privacy and location leaks in LTE \cite{jover2016lte}.

The LTE protocol specifications also define vulnerable edge cases that, despite being rarely executed, are still supported by the protocol. For example, although it is very unlikely that a UE would ever transmit an \emph{Attach Request} message using its IMSI as the identifier, the protocol describes specific scenarios in which this would occur. For example, during network recovery after the core network lost the UE's temporary identifiers. In this case, the network can trigger the mobile device to retransmit the \emph{Attach Request} message with its IMSI in the clear 
\cite{LTE_NAS}. 

In a nutshell, most active LTE protocol exploits occur because of a combination of the protocol supporting insecure edge cases and the implicit trust of pre-authentication messages~\cite{jover2016exploits}. The first two columns of Table III 
summarize some of the most relevant LTE protocol exploits that have been identified in open literature in the recent past. 

\subsection{Impact on 5G Networks}
Most of the known LTE protocol security vulnerabilities were studied and dissected by the security working group of 3GPP \cite{LTE_SECURITY_STUDY} with the aim for defining a secure 5G standard. As a result of that study, specific security goals for 5G mobile networks were set to address the problem of IMSI catchers, pre-authentication messages and location leaks. Device and user tracking leveraging the Radio Network Temporary Identifier (RNTI) \cite{jover2016lte} was, on the other hand, disregarded in the 3GPP study because RNTIs are claimed to be short lived identifiers that cannot be leveraged for privacy leaks. Recent research, however, confirmed that the RNTI can be used to track subscribers \cite{rupprecht-19-layer-two}.

As discussed in this paper, despite 
being highly sophisticated and robust against adversarial attacks, the 5G security framework still includes a number of edge cases that facilitate bypassing all security functions. Hence, most of the demonstrated LTE protocol exploits are not fully addressed and are still a potential threat, as described in the third column of Table III. 

These findings put pressure on 5G. Unlike in the case of LTE, where most security research and resulting protocol weaknesses were identified after the protocol was defined, implemented and globally deployed, the security research community is moving fast with 5G. Weaknesses in the 5G specifications are being identified as the specifications are released \cite{5GAKAformalanalysis2018}.

Note that most deployed LTE networks still rely on many early 3GPP Release 8 or 9 features. This highlights that once deployed, after years of standardization, certification and testing, major network upgrades can take considerable time to be widely implemented, for a variety of reasons. Since security cannot be considered an add-on feature, the advantage of providing early awareness of potential security issues is that these can be pragmatically analyzed, fixed, and the specifications revised during the initial roll outs and before mass commercial deployments of networks, UEs, and services. It is of critical importance that the lessons learned from LTE are applied now to design a 5G system architecture that is fully resilient to protocol exploits. In other words, 5G security should be addressed in the current Release 15 as opposed to in upcoming releases.


\section{Conclusions}
\label{sec:conclusions}
Wireless communication security has always been of critical importance 
and will be more so as technology evolves towards 5G. Traditionally used mostly for non-critical voice communication, many of the current and emerging data and control communication systems that leverage cellular access networks have stringent requirements in terms of integrity and privacy of user data. Applications include tactical communication, first responder ad-hoc networks, and mission-critical IoT.

This paper provides the first holistic analysis of the first version of the 5G security specifications \cite{5G_SECURITY_3GPP}. Our study highlights a number of potential insecure protocol edge cases and limitations that result from infeasible requirements or assumptions. Despite clearly targeting to address the known security vulnerabilities of LTE networks, the 5G specifications are, as of Release 15, Version 1.0.0, still vulnerable to the same types of LTE adversarial attacks that leverage pre-authentication messages.

Global adoption and enforcement of a robust security framework is necessary to avoid having to support insecure operational modes and rely on implicit trust of pre-authentication messages. Therefore it is critical to ensure that no insecure edge cases are supported by the 5G standard. In particular, null authentication, null encryption, downgrade attacks, exploitation of pre-authentication messages, and SUPI catching shall not be facilitated in any mode of 5G network operation. The success of such a security framework should not be subject to implicit assumptions 
or implementation options neither. 

While the 5G security architecture made a substantial leap in the right direction with the proposed PKI architecture, security research and development is still necessary to fully address the known and new security vulnerabilities of next generation mobile communication systems. Standardization bodies, researchers, regulators, and industry all need to work together to accomplish a securer architecture, design, development and deployment of emerging and future mobile communication and control systems. Global cooperation and collaboration, led by the standardization bodies, is necessary to define and implement the required system architecture and CA that would provide the foundation for secure 5G systems.

\balance

\bibliographystyle{IEEEtran}
\bibliography{imsi_bib,vuk}

\begin{thebibliography}{10}
\providecommand{\url}[1]{#1}
\csname url@samestyle\endcsname
\providecommand{\newblock}{\relax}
\providecommand{\bibinfo}[2]{#2}
\providecommand{\BIBentrySTDinterwordspacing}{\spaceskip=0pt\relax}
\providecommand{\BIBentryALTinterwordstretchfactor}{4}
\providecommand{\BIBentryALTinterwordspacing}{\spaceskip=\fontdimen2\font plus
\BIBentryALTinterwordstretchfactor\fontdimen3\font minus
  \fontdimen4\font\relax}
\providecommand{\BIBforeignlanguage}[2]{{%
\expandafter\ifx\csname l@#1\endcsname\relax
\typeout{** WARNING: IEEEtran.bst: No hyphenation pattern has been}%
\typeout{** loaded for the language `#1'. Using the pattern for}%
\typeout{** the default language instead.}%
\else
\language=\csname l@#1\endcsname
\fi
#2}}
\providecommand{\BIBdecl}{\relax}
\BIBdecl

\bibitem{swindlehurst2014millimeter}
A.~L. Swindlehurst, E.~Ayanoglu, P.~Heydari, and F.~Capolino, ``Millimeter-wave
  massive mimo: The next wireless revolution?'' \emph{IEEE Communications
  Magazine}, vol.~52, no.~9, pp. 56--62, 2014.

\bibitem{docomo2014docomo}
N.~Docomo, ``{DOCOMO 5G} white paper, {5G} radio access: Requirements, concept
  and technologies,'' \emph{White Paper, Jul}, 2014.

\bibitem{boccardi2014five}
F.~Boccardi, R.~W. Heath, A.~Lozano, T.~L. Marzetta, and P.~Popovski, ``Five
  disruptive technology directions for {5G},'' \emph{IEEE Communications
  Magazine}, vol.~52, no.~2, pp. 74--80, 2014.

\bibitem{LTE_SECURITY_3GPP}
{3rd Generation Partnership Project (3GPP), Technical Specification Group
  Services and System Aspects}, ``{3GPP system architecture evolution (SAE) -
  security architecture (Release 14)},'' \emph{3GPP TS 33.401, V14.5.0}, Jan.
  2018.

\bibitem{rupprecht-19-layer-two}
D.~Rupprecht, K.~Kohls, T.~Holz, and C.~P\"{o}pper, ``Breaking {LTE} on layer
  two,'' in \emph{IEEE Symposium on Security \& Privacy (SP)}.\hskip 1em plus
  0.5em minus 0.4em\relax IEEE, May 2019.

\bibitem{jover2016exploits}
\BIBentryALTinterwordspacing
R.~P. Jover, ``{LTE} security, protocol exploits and location tracking
  experimentation with low-cost software radio,'' \emph{CoRR}, vol.
  abs/1607.05171, 2016. [Online]. Available:
  \url{http://arxiv.org/abs/1607.05171}
\BIBentrySTDinterwordspacing

\bibitem{LTEpracticalattacks}
A.~Shaik, R.~Borgaonkar, N.~Asokan, V.~Niemi, and J.-P. Seifert, ``Practical
  attacks against privacy and availability in {4G/LTE} mobile communication
  systems,'' in \emph{{Proceedings of the 23rd Annual Network and Distributed
  System Security Symposium (NDSS 2016)}}, 2016.

\bibitem{jover2016opensource}
R.~P. Jover, ``{The impact of open source on mobile security research},'' Tech.
  Rep., May 2016, \url{https://goo.gl/hi4ukn}.

\bibitem{RaoMILCOM17}
\BIBentryALTinterwordspacing
R.~M. Rao, V.~Marojevic, S.~Ha, and J.~Reed, ``{LTE PHY} layer vulnerability
  analysis and testing using open-source {SDR} tools,'' in \emph{Proc. IEEE
  MILCOM2017}, "Baltimore, MD, USA", 23-25 October 2017. [Online]. Available:
  \url{http://dx.doi.org/10.1109/MILCOM.2017.8170787}
\BIBentrySTDinterwordspacing

\bibitem{hussain2018lteinspector}
S.~R. Hussain, O.~Chowdhury, S.~Mehnaz, and E.~Bertino, ``{LTEInspector}: A
  systematic approach for adversarial testing of {4G LTE},'' in \emph{Symposium
  on Network and Distributed Systems Security (NDSS)}, 2018, pp. 18--21.

\bibitem{IMSIcatcher}
D.~Strobel, ``{IMSI catcher},'' \emph{Chair for Communication Security,
  Ruhr-Universit{\"a}t Bochum}, p.~14, 2007.

\bibitem{5G_SECURITY_3GPP}
{3GPP Technical Specification Group Services and System Aspects}, ``{Security
  architecture and procedures for 5G system},'' \emph{3GPP TS 33.501, V1.0.0},
  March 2018.

\bibitem{LTE_SECURITY_STUDY}
------, ``{Study on the security aspects of the next generation system (Release
  14)},'' \emph{3GPP TR 33.899 V1.3.0}, August 2017.

\bibitem{LabCM2017}
M.~Labib, V.~Marojevic, J.~H. Reed, and A.~I. Zaghloul, ``Enhancing the
  robustness of {LTE} systems: Analysis and evolution of the cell selection
  process,'' \emph{IEEE Commun. Mag.}, vol.~55, no.~2, pp. 208--215, Feb. 2017.

\bibitem{HuaweiBan}
``{Pentagon Orders Stores on Military Bases to Remove Huawei, ZTE Phones},''
  {The Wall Street Journal}, May 2018, \url{https://goo.gl/ciySYB}.

\bibitem{jover2016lte}
R.~P. Jover, ``{LTE security and protocol exploits},'' \emph{{Shmoocon 2016}},
  January 2016.

\bibitem{LabSM2017}
\BIBentryALTinterwordspacing
M.~Labib, V.~Marojevic, J.~H. Reed, and A.~I. Zaghloul, ``Extending {LTE} into
  the unlicensed spectrum: Technical analysis of the proposed variants,''
  \emph{IEEE Communications Standards Magazine}, vol.~1, no.~4, pp. 31--39,
  December 2017. [Online]. Available:
  \url{http://dx.doi.org/10.1109/MCOMSTD.2017.1700040}
\BIBentrySTDinterwordspacing

\bibitem{5GAKAformalanalysis2018}
D.~{Basin}, J.~{Dreier}, L.~{Hirschi}, S.~{Radomirovi{\'c}}, R.~{Sasse}, and
  V.~{Stettler}, ``Formal analysis of {5G} authentication,'' \emph{ArXiv
  e-prints}, jun 2018.

\bibitem{5GAKAsecurity}
M.~Dehnel-Wild and C.~Cremers, ``Security vulnerability in {5G-AKA} draft,''
  \emph{Department of Computer Science, University of Oxford, Tech. Rep}, 2018.

\bibitem{AKAkoutsos20185g}
A.~Koutsos, ``The {5G-AKA} authentication protocol privacy,'' \emph{arXiv
  preprint arXiv:1811.06922}, 2018.

\bibitem{5Gdowngrade}
M.~Khan, P.~Ginzboorg, K.~J{\"a}rvinen, and V.~Niemi, ``Defeating the downgrade
  attack on identity privacy in {5G},'' \emph{arXiv preprint arXiv:1811.02293},
  2018.

\bibitem{housley1998internet}
R.~Housley, W.~Ford, W.~Polk, and D.~Solo, ``Internet x. 509 public key
  infrastructure certificate and crl profile,'' Tech. Rep., 1998.

\bibitem{SSL}
A.~Freier, P.~Karlton, and P.~Kocher, ``The secure sockets layer (ssl) protocol
  version 3.0,'' Tech. Rep., 2011.

\bibitem{KambourakisPKI}
\BIBentryALTinterwordspacing
G.~Kambourakis, A.~Rouskas, and S.~Gritzalis, ``Performance evaluation of
  public key-based authentication in future mobile communication systems,''
  \emph{EURASIP J. Wirel. Commun. Netw.}, vol. 2004, no.~1, pp. 184--197, Aug.
  2004. [Online]. Available: \url{http://dx.doi.org/10.1155/S1687147204403016}
\BIBentrySTDinterwordspacing

\bibitem{5GPPPcerts}
P.~Bisson and J.~Waryet, ``{5GPPP} phase 1 security landscape,'' \emph{5G PPP
  Security Group White Paper}, 2017.

\bibitem{attack_sniffing}
K.~Nohl and S.~Munaut, ``Wideband {GSM} sniffing,'' in \emph{In 27th Chaos
  Communication Congress}, 2010, \url{http://goo.gl/wT5tz}.

\bibitem{LTEpracticalattacks_Blackhat}
A.~Shaik, R.~Borgaonkar, N.~Asokan, V.~Niemi, and J.-P. Seifert, ``{LTE} and
  {IMSI} catcher myths,'' \emph{Proc. of BlackHat Europe}, 2015.

\bibitem{MarVTC17}
\BIBentryALTinterwordspacing
V.~Marojevic, R.~M. Rao, S.~Ha, and J.~Reed, ``Performance analysis of a
  mission-critical portable {LTE} system in targeted {RF} interference,'' in
  \emph{Proc. IEEE Int. Veh. Tech. Conf. (VTC)}, Toronto, Canada, 23-27
  September 2017. [Online]. Available:
  \url{http://dx.doi.org/10.1109/VTCFall.2017.8288187}
\BIBentrySTDinterwordspacing

\bibitem{LabICNC2015}
M.~Labib, V.~Marojevic, J.~H. Reed, and A.~I. Zaghloul, ``How to enhance the
  immunity of {LTE} systems against {RF} spoofing,'' in \emph{Proc. 2016 Int.
  Conf. Computing, Networking and Communications (ICNC)}, Kauai, HI, USA, 15-18
  February 2016, pp. 1--5.

\bibitem{MINA_JP}
M.~Labib, V.~Marojevic, and J.~Reed, ``Analyzing and enhancing the resilience
  of {LTE/LTE-A},'' in \emph{Proc. IEEE Conf. Standards for Communications and
  Networking (CSCN)}, Tokyo, Japan, October 2015.

\bibitem{Nor2016}
K.~Norrman, M.~N{\"a}slund, and E.~Dubrova, ``Protecting imsi and user privacy
  in {5G} networks,'' in \emph{Proceedings of the 9th EAI International
  Conference on Mobile Multimedia Communications}.\hskip 1em plus 0.5em minus
  0.4em\relax ICST (Institute for Computer Sciences, Social-Informatics and
  Telecommunications Engineering), 2016, pp. 159--166.

\bibitem{LTE_NAS}
{Universal Mobile Telecommunications System (UMTS) - LTE},
  ``{Non-Access-Stratum (NAS) protocol for Evolved Packet System (EPS) - Stage
  3},'' \emph{{3GPP TS 24.301, V9.11.0}}, 2013.

\end{thebibliography}

\end{document}